\documentclass[aps,prd,preprint,showpacs,superscriptaddress,groupedaddress,nofootinbib]{revtex4-1}
\usepackage{epsfig}
\usepackage{graphicx}
\usepackage{color}
\usepackage[normalem]{ulem}
\usepackage{amsmath}
\usepackage[breaklinks, plainpages=false, colorlinks=true, anchorcolor=cyan, linkcolor=red, citecolor=cyan, urlcolor=magenta, bookmarks=false]{hyperref}
\usepackage[caption=false]{subfig}
\begin{document}

\renewcommand{\thefigure}{\arabic{figure}}
\setcounter{figure}{0}

\def \D {{\cal D}}
\def \univ{{\bf{n}}}
\def \unp{{\bf{p}}}
\def \uk{{\bf{k}}}
\def \uniD{{\boldsymbol{\Delta}}}
\def \uniP{{\boldsymbol{\delta}}}

\bibliographystyle{apsrev}

\title{Time-Delay Interferometry and Clock-Noise Calibration} 
\author{Massimo Tinto} 
\email{mtinto@ucsd.edu}
\affiliation{University of California San Diego,\\
  Center for Astrophysics and Space Sciences, \\
  9500 Gilman Dr, La Jolla, CA 92093, \\
  U.S.A.}  
\author{Olaf Hartwig}
\email{olaf.hartwig@aei.mpg.de}
\affiliation{Max-Planck-Institut f\"ur Gravitationsphysik
  (Albert-Einstein-Institut), Callinstra{\ss}e 38, 30167 Hannover,
  Germany}

\date{\today}

\begin{abstract}
  The Laser Interferometer Space Antenna is a joint ESA-NASA
  space-mission to detect and study mHz cosmic gravitational waves.
  The trajectories followed by its three spacecraft result in unequal-
  and time-varying arms, requiring use of the Time-Delay
  Interferometry (TDI) post-processing technique to cancel the laser
  phase noises affecting the heterodyne one-way Doppler
  measurements. Although the second-generation formulation of TDI
  cancels the laser phase noises when the array is both rotating and
  ``flexing'', second-generation TDI combinations for which the phase
  fluctuations of the onboard ultra stable oscillators (USOs) can be
  calibrated out have not appeared yet in the literature. In this
  article we present the solution of this problem by generalizing to
  the realistic LISA trajectory the USO calibration algorithm derived
  by Armstrong, Estabrook and Tinto for a static configuration.
\end{abstract}

\maketitle

\section{Introduction}
\label{SECI}

Gravitational waves (GWs) are predicted by Einstein's theory of
general relativity and represent disturbances of space-time
propagating at the speed of light. Because of their extremely small
amplitudes and interaction cross-sections, GWs carry information about
regions of the Universe that would be otherwise unobtainable through
the electromagnetic spectrum.  Their first detection announced by the
LIGO project in February 2016 \cite{GW150914}, followed by the
additional observations of four more events \cite{LV2,LV3,LV4,LV5},
mark the beginning of GW Astronomy.

Contrary to ground-based laser interferometers, which are sensitive to
GWs in a band from about a few tens of Hz to a few kilohertz,
space-based interferometers are expected to access the frequency
region from a few tenths of millihertz to about a few tens of Hz,
where GW signals are expected to be larger in number and characterized
by larger amplitudes. The most notable example of a space-based
interferometer, which for several decades has been jointly studied in
Europe and in the United States of America, is the Laser
Interferometer Space Antenna (LISA) mission~\cite{LISA2017}.  LISA,
which is now expected to be launched in the year 2034, will detect and
study cosmic gravitational waves in the $10^{-4} - 1$ Hz band by
relying on coherent laser beams exchanged by three remote
spacecraft along the arms of their forming giant (almost) equilateral
triangle of $2.5 \times 10^6$ km arm-length.

A space-based laser interferometer GW detector measures relative
frequency changes experienced by coherent laser beams exchanged by
three pairs of spacecraft.  As the laser beams are received, they are
made to interfere with the outgoing laser light. Since the received
and receiving frequencies of the laser beams can be different by tens
to perhaps hundreds of MHz (consequence of the Doppler effect from the
relative inter-spacecraft velocities and the intrinsic frequency
differences of the lasers), to remove these large beat-notes present
in the heterodyne measurements one relies on the use of a microwave
signal generated by an onboard clock (usually referred to as
Ultra-Stable Oscillator (USO)).  The magnitude of the frequency
fluctuations introduced by the USO into a heterodyne one-way Doppler
measurement depends linearly on the USO's noise itself and the
heterodyne beat-note frequency. Space-qualified, state of the art
clocks are oven-stabilized crystals characterized by an Allan standard
deviation of ${\sigma_A} \approx 10^{-13}$ for averaging times in the
interval $1 - 10^4$ s, covering the frequency band of interest to
space-based interferometers~\cite{LISA2017}. In the case of the LISA
mission, in particular, it was estimated~\cite{TEA02} that the
magnitude of the square-root of the power spectral density of the
USO's relative frequency fluctuations appearing, for instance, in the
unequal-arm Michelson TDI combination $X$ (valid for a static-array
configuration) would be about three orders of magnitude larger than
that due to the residual (optical path and proof-mass) noise sources.

A technique for removing the USO noise from a Michelson interferometer
for a static array configuration was first discussed in
\cite{Hellingsetal}, applied in \cite{Hellings} to the unequal-arm
Michelson $X$ and Sagnac $\alpha$ TDI combinations for a static array
(TDI-1) and improved and extended to all the TDI-1 combinations in
\cite{TEA02}. This technique requires the modulation of the laser
beams exchanged by the spacecraft, and the further measurement of six
more inter-spacecraft relative phases by comparing the sidebands of
the received beam against sidebands of the transmitted beam. The
physical reason behind the use of modulated beams for calibrating the
USOs noises is to exchange the USOs phase fluctuations with the same
time delays as those experienced by the laser phase noises as they
propagate along the three arms. By performing sideband-sideband
measurements~\cite{Hellingsetal, Hellings, TEA02} six additional
one-way phase differences are generated that allow one to calibrate
out the USOs phase fluctuations from the TDI-1 combinations while
preserving the gravitational wave signal in the resulting
USO-calibrated TDI data.

Although an alternative experimental implementation to the modulation
technique has recently been proposed \cite{TintoYu2015}, which relies
on the use of an onboard optical-frequency comb \cite{Hall, Hansch,
  Holzwarth} to generate the microwave frequency coherent to the
frequency of the onboard laser \footnote{The optical frequency-comb
  technique exactly cancels the microwave signal phase fluctuations as
  it relies on modified TDI-2 combinations and does not require use of
  modulated beams.}, in this article we derive the TDI combinations
for a rotating and ``flexing'' array (so called second-generation TDI,
TDI-2) that calibrate out the microwave signal phase fluctuations due
to the onboard LISA USOs. A summary of this article is presented
below.

In Section~\ref{SECII} we provide the mathematical expressions
describing the one-way heterodyne measurements performed onboard the
LISA spacecraft. They reflect the planned LISA's split-interferometry
design, and they were first presented in
\cite{OttoHeinzelDanzmann}. There Otto {\it et al} also proposed a
data processing algorithm to obtain TDI-2 combinations that are USO
noise-free. Their approach, however, has recently been shown (by the
LISA Simulation Working Group) to not work as expected and we discuss
the physical reasons behind its short-coming.

After showing that the commutator of two delay-operators applied to
the phase noise of a LISA USO results in relative frequency
fluctuations (strain) that are significantly smaller than those
associated with the acceleration and optical-path noises, in
Section~\ref{SECIII} we derive the expressions that calibrate the USO
noises out of the TDI-2 unequal-arm Michelson and Sagnac
interferometric combinations \cite{TD2014}. A summary of our results
and conclusions are then presented in Section~\ref{SECIV}.

\section{Split-Interferometry One-Way Heterodyne Measurements}
\label{SECII}

In what follows we provide the expressions for the eight one-way
heterodyne measurements performed onboard spacecraft \# $1$; the
remaining $16$ can be obtained by cyclic permutation of the spacecraft
indices. These expressions were discussed in
\cite{OttoHeinzelDanzmann} in the context of the LISA
split-interferometry architecture and we refer the reader to that
publication for more details. Here we modify those expressions by
multiplying both sides of the equations given there by the functions
$\theta_{ij}$, which are either equal to $+ 1$ or $- 1$ depending on
whether the frequency of the received beam, $\nu_i$, is larger or
smaller than the frequency $\nu_j$ of the receiving beam
respectively. This results in modified expressions for the fractional
frequency beat-note coefficients, which are given below.

\begin{figure}
\centering
\includegraphics[width=5in]{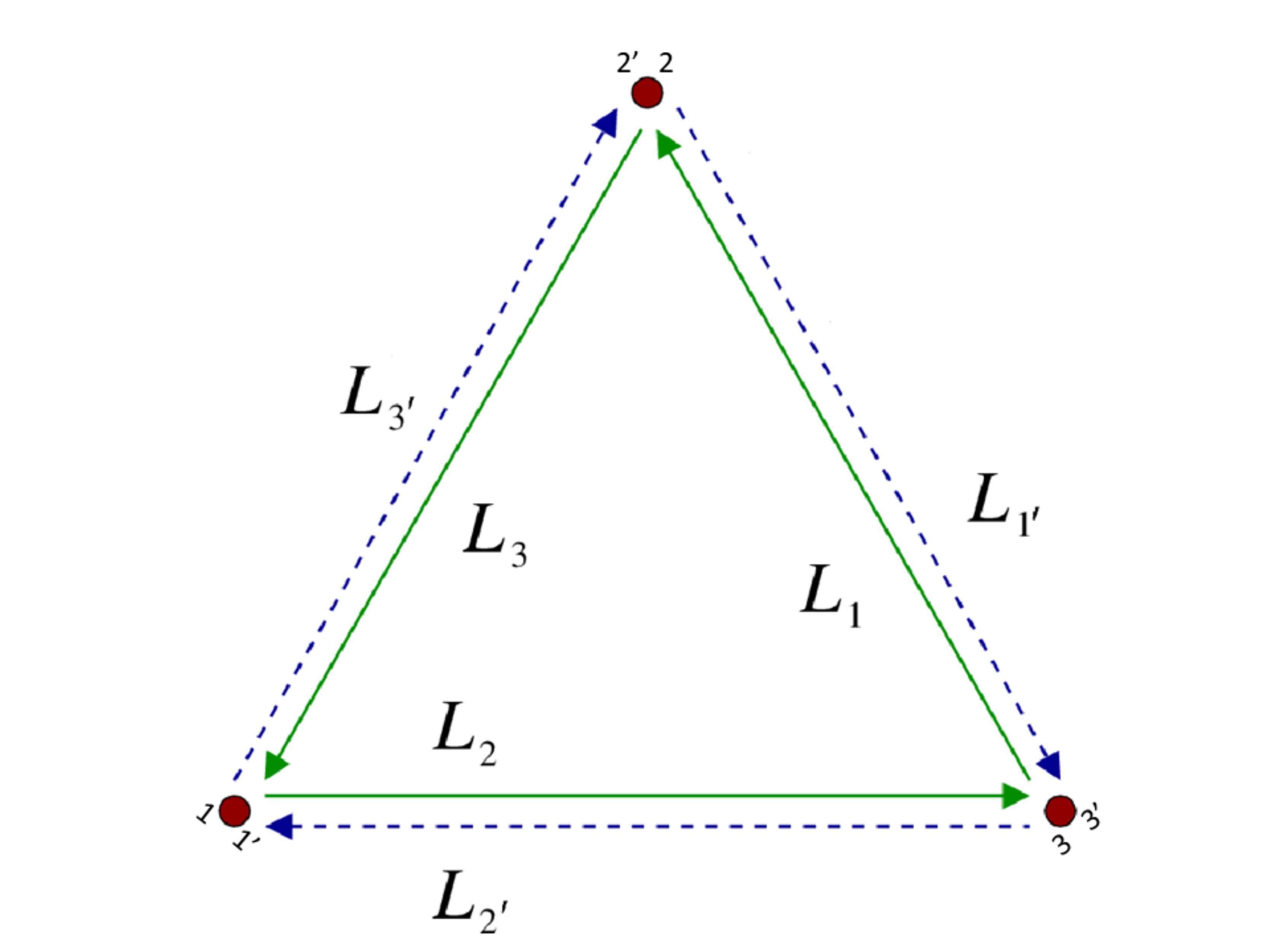}
\caption{\label{fig1} Schematic LISA configuration. The spacecraft are
  labeled $1$, $2$, and $3$, with their optical benches labeled with
  primed or unprimed indices depending on whether the received laser
  beam is propagating clockwise or counter-clockwise as seen from
  above the plane of the picture. The optical paths are denoted by
  $L_i$, $L_{i'}$ where the index $i$ corresponds to the opposite
  spacecraft.}
\end{figure}

We adopt the description of the LISA array given in \cite{TD2014}, in
which the beam arriving at spacecraft $i$ has subscript $i$ and is
primed or unprimed depending on whether the beam is traveling
respectively clockwise or counter-clockwise around the LISA triangle
as seen from above the plane of the constellation described in
Fig. (\ref{fig1}).  We also adopt the usual notation for delayed data
streams, which is convenient for algebraic manipulations
\cite{TD2014}. We define the six time-delay operators $\D_{i}$,
$\D_{i'}$, $i = 1, 2, 3 \ , \ i' = 1', 2', 3'$, where for any data
stream $x(t)$
\begin{equation}
  \D_{j'} \D_{i} x(t) = x(t - L_i(t - L_{j'}) - L_{j}),
\end{equation}
where ($L_{j'}, L_{i}$), $j' = 1', 2', 3' \ , \ i = 1, 2, 3$, are the
light travel times along the three arms of the LISA triangle along the
clock-wise and anti-clock-wise directions respectively \footnote{The
  speed of light $c$ is assumed to be unity in this article}. It is
important to note that, although $[\D_{j'} \ , \ \D_{i}] x(t) \ne 0$
in general, the commutator of two delay operators can be regarded as
equal to zero if the resulting magnitude of a random process it is
applied to is significantly smaller than the magnitude of the
secondary (acceleration and optical-path) noises affecting the LISA
measurements.

\begin{figure}
\centering
\includegraphics[width=5in]{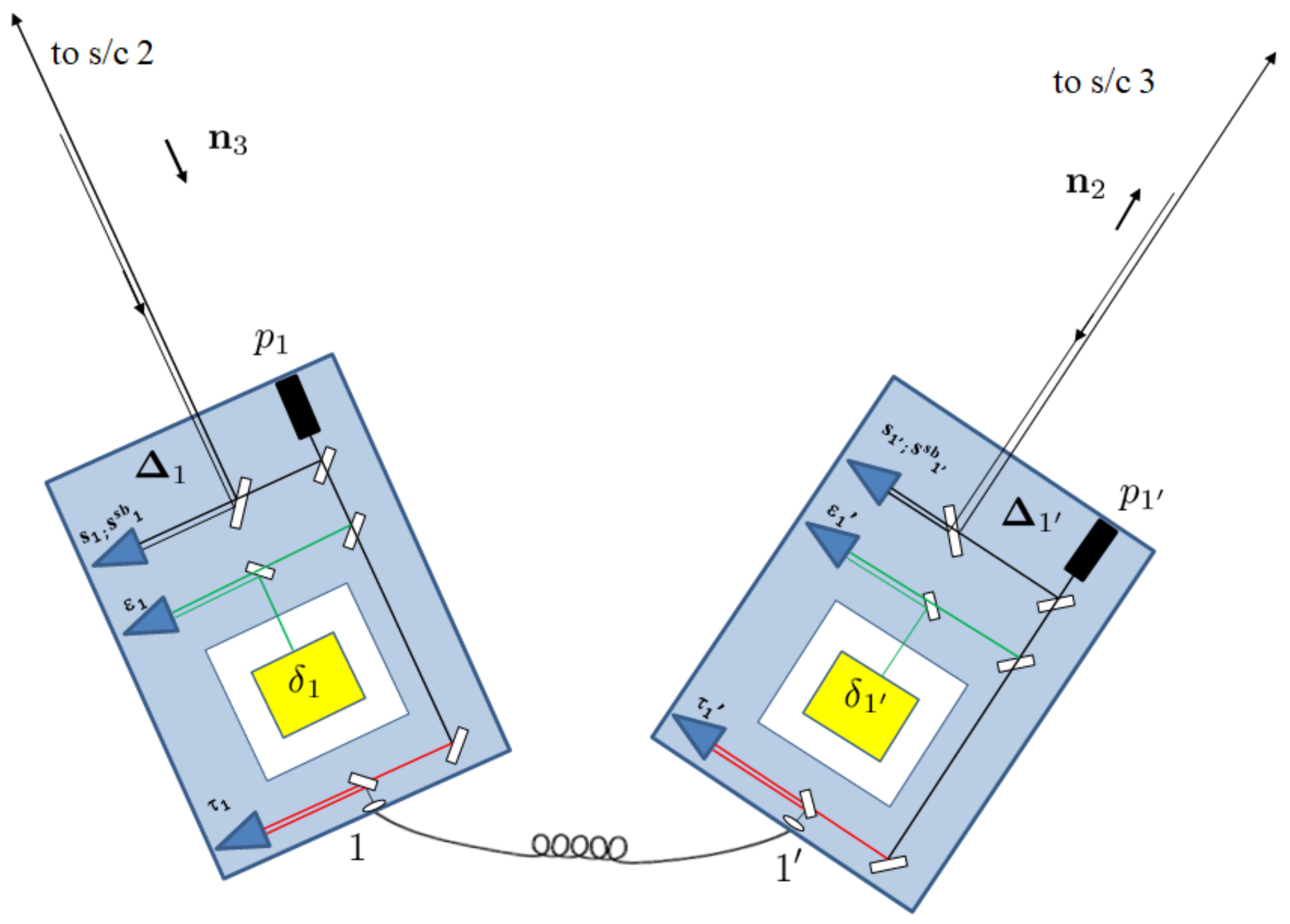}
\caption{\label{fig2} Schematic diagram of the proof-mass and optical
  bench assemblies for LISA spacecraft \# 1. The left bench reads out
  a phase signal $s^{\rm c}_{1}$ from optical bench $2'$ onboard
  spacecraft \# $2$. The phase difference is measured by using the
  laser and the photo-detector on the left optical bench. The motion
  of the optical bench relative to the proof-mass is measured through
  internal metrology and results in the time series
  $\epsilon_{1}$. The relative phase fluctuations between the laser on
  the optical bench $1$ and the laser on the optical bench $1'$ are
  instead captured by the measurements $\tau_{1}$ and $\tau_{1'}$
  respectively. Finally, the sideband-sideband phase differences,
  $s^{\rm sb}_{1}$ and $s^{\rm sb}_{1'}$, capture the phase
  fluctuations of the USOs onboard spacecraft \# $2$ and $3$
  respectively relative to those of the USO onboard spacecraft
  $1$. These are the ``essential ingredients'' of the USO calibration
  algorithm.}
\end{figure}

The eight heterodyne measurements performed onboard spacecraft \# $1$
are presented in two groups, each including four data set from the
specific optical bench where they are collected. The group of
measurements from optical bench $1'$ are represented by the following
mathematical expressions (see Fig.(\ref{fig2}))

\begin{eqnarray}
s^{\rm c}_{1'} & = & \left[H_{1'} + \D_{2'} p_{3} - p_{1'} -
  2 \pi \nu_{3} (\univ_{2'} \cdot \D_{2'} \uniD_{3} + \univ_{2}
  \cdot \uniD_{1'}) + N_{1'}^{\rm obt} \right] - a_{1'} q_1 +
N_{1'}^{\rm s} \ ,
\label{s1p}
\\
s^{\rm sb}_{1'} & = & \left[H_{1'} + \D_{2'} p_{3} - p_{1'} +
  m_3 \D_{2'} q_{3} - m_{1'} q_1 - 
  2 \pi \nu_{3} (\univ_{2'} \cdot \D_{2'} \uniD_{3} + \univ_{2}
  \cdot \uniD_{1'}) + N_{1'}^{\rm obt.sb} \right] 
\nonumber
\\
& - & c_{1'} q_1 + N_{1'}^{\rm sb} \ ,
\label{s1psb}
\\
\epsilon_{1'} & = & \left[p_{1} - p_{1'} +
  4 \pi \nu_{1} (\univ_{2} \cdot \uniP_{1'} - \univ_{2}
  \cdot \uniD_{1'}) + \mu_1 \right] - b_{1'} q_1 + N_{1'}^{\epsilon} \ ,
\label{epsp}
\\
\tau_{1'} & = & \left[p_{1} - p_{1'} +
  \mu_1 \right] - b_{1'} q_1 + N_{1'}^{\tau} \ ,
\label{taup}
\end{eqnarray}
while those from optical bench $1$ are equal to
\begin{eqnarray}
s^{\rm c}_{1} & = & \left[H_{1} + \D_{3} p_{2'} - p_{1} -
  2 \pi \nu_{2'} (\univ_{3} \cdot \D_{3} \uniD_{2'} + \univ_{3'}
  \cdot \uniD_{1}) + N_{1}^{\rm obt} \right] - a_{1} q_1 +
N_{1}^{\rm s} \ ,
\label{s1}
\\
s^{\rm sb}_{1} & = & \left[H_{1} + \D_3 p_{2'} - p_{1} +
  m_{2'} \D_{3} q_{2} - m_{1} q_1 - 
  2 \pi \nu_{2'} (\univ_{3} \cdot \D_{3} \uniD_{2'} + \univ_{3'}
  \cdot \uniD_{1}) + N_{1}^{\rm obt.sb} \right] 
\nonumber
\\
& - & c_{1} q_1 + N_{1}^{\rm sb} \ ,
\label{s1sb}
\\
\epsilon_{1} & = & \left[p_{1'} - p_{1} +
  4 \pi \nu_{1'} (\univ_{3'} \cdot \uniP_{1} - \univ_{3'}
  \cdot \uniD_{1}) + \mu_{1'} \right] - b_{1} q_1 + N_{1}^{\epsilon} \ ,
\label{eps}
\\
\tau_{1} & = & \left[p_{1'} - p_{1} +
  \mu_{1'} \right] - b_{1} q_1 + N_{1}^{\tau} \ .
\label{tau}
\end{eqnarray}
The observables $s^{\rm c}, s^{\rm sb}, \epsilon, \tau$ are the
inter-spacecraft carrier-to-carrier and sideband-to-sideband one-way
heterodyne measurements, the proof-mass-to-optical bench and the
bench-to-bench metrology measurements respectively; the $a, b, c$ are
the fractional frequency beat-note coefficients, i.e. the coefficients
determined by the phasemeter \cite{OttoHeinzelDanzmann} that multiply
the USO-referenced pilot-tone frequency so as to match the beat-note
frequencies. The expressions of these coefficients are obtained from
the corresponding ones given in \cite{OttoHeinzelDanzmann} by
multiplying them by the appropriate $\theta$s. After performing such
operation it is easy to show that they become equal to
\begin{eqnarray}
a_{1'} & = & \frac{\nu_3(1 - {\dot L}_{2'}) - \nu_{1'}}{f_1} \ \ , 
\label{ap}
\\
a_{1} & = &  \frac{\nu_{2'}(1 - {\dot L}_{3}) - \nu_{1}}{f_1} 
\label{a}
\\
b_{1'} & = & \frac{\nu_1 - \nu_{1'}}{f_1} \ \ , \ \ 
b_{1} = \frac{\nu_{1'} - \nu_{1}}{f_1} = - b_{1'}
\label{b}
\\
c_{1'} & = & \frac{(\nu_3 + m_3 f_3)(1 - {\dot L}_{2'}) - (\nu_{1'} +  m_{1'} f_1)}{f_1} \ \ , 
\label{cp}
\\
c_{1} & = & \frac{(\nu_{2'} + m_{2'} f_2)(1 - {\dot L}_{3}) - (\nu_{1} +  m_{1} f_1)}{f_1}
\label{c}
\end{eqnarray}

In equations (\ref{s1p} - \ref{tau}) the $H$-terms are the
contributions to the measured phase fluctuations due to a possibly
present transverse-traceless gravitational wave signal; the $p$- and
$\nu$-terms represent the lasers' phase noises and frequencies
respectively; the $q$-terms are the phase noises due to the three
USOs; the $N$-terms are shot-noise phase fluctuations at the
photo-detectors; $L$-terms and $\dot L$-terms are the inter-spacecraft
relative optical paths and their time-derivatives respectively; the
$f$-terms are the USOs' pilot-tones frequencies while the $m$-terms are
integer numbers defining the modulation frequencies
\cite{OttoHeinzelDanzmann}; the $\univ$-terms are unit vectors along
the directions of propagation of the laser beams; the $\uniD$-terms
and $\uniP$-terms are vector random processes associated with the
mechanical vibrations of the optical benches and proof-masses with
respect to the local inertial reference frame respectively; the
$\mu$-terms are phase fluctuations due to the optical fibers linking
the two optical benches and they have been assumed to be independent
of the direction of propagation of the optical beams within them (see
\cite{OttoHeinzelDanzmann} for a clear discussion about this point);
finally the $\D_i$, $\D_{j'}$ are delay operators \cite{TD2014}.

Since the LISA array is both rotating and ``flexing'', two delay
operators do not commute in general \cite{TD2014}. For instance, with
a laser noise equal to $30$ Hz/${\sqrt {\rm Hz}}$ in the mHz band
\cite{TD2014}, the commutator of two delay operators applied to it
results in residual fluctuations that are about an order of magnitude
larger than those identified by the secondary noises. This is in fact
the reason why the formulation of TDI for a static array is unable to
suppress such a laser noise below the level identified by the
secondary noises.

In the case of a LISA's USO, however, its relative frequency
fluctuations at optical frequency are significantly smaller than those
of a laser. This means that the commutator of two delay operators
applied to it results in relative frequency fluctuations significantly
smaller than those due to the secondary noises. To be more
quantitative, let us estimate the magnitude of the commutator of two
delay operators applied to the phase fluctuations, $q(t)$, of a LISA
USO. This is given by the following expression
\begin{equation}
[\D_i, \D_j]q(t) = q(t - L_i (t) - L_j(t - L_i)) - q(t - L_j(t) - L_i(t
- L_j)) \simeq [L_i {\dot L}_j - L_j {\dot L}_i] {\dot q} \ .
\label{commut}
\end{equation}
The right-hand-side of equation (\ref{commut}) implies the following
order-of-magnitude estimate of the corresponding Fourier components of
the relative frequency fluctuations (strain) amplitude,
$\frac{| \widetilde{\Delta C_q} (f) |}{\nu_0}$, in a TDI combination
\begin{equation}
\frac{| \widetilde{\Delta C_q} (f)|}{\nu_0} \equiv \left( 4 \pi L
  {\dot L} \right) \left( \frac{a f_q}{\nu_0} \right)
\left(\frac{|{\widetilde{\dot q}} (f)|}{2 \pi f_q} \right) \ f \ ,
\label{Cq}
\end{equation}
where the $\widetilde{}$ symbol means ``Fourier transform'', and $f$
is the Fourier frequency. By assuming a LISA USO characterized by a
one-sided power spectral density of relative frequency fluctuations
equal to $S_y(f) = 6.7 \times 10^{-27} f^{-1} \ {\rm Hz}^{-1}$
\footnote{This one-sided power spectral density corresponds to an
  Allan standard deviation equal to about $10^{-13}$ from $1$ to
  $10^4$ seconds integration times \cite{Barnes_etal}.}, a beat-note
frequency $a f_q = 25 \ {\rm MHz}$, a laser frequency
$\nu_0 = 3 \times 10^{14}$ Hz, a LISA arm-length (in seconds)
$L = 8.3$ sec. and an inter-spacecraft characteristic relative
velocity $\dot L \simeq 3 \times 10^{-8}$, we find the right-hand-side
of equation (\ref{Cq}) to be equal to
$2.3 \times 10^{-26} \ {\rm Hz}^{-1/2}$ at $f = 1 \ {\rm Hz}$ and
smaller than this value at lower frequencies. Since this is more than
five orders of magnitude smaller than the minimum of the strain
sensitivity identified by the secondary noises in the TDI-1
combinations \cite{TEA02}, we can treat two delay operators as
commuting when they are applied to a LISA USO phase noise.

Before deriving the USO noise calibrating expressions, it is necessary
to first remove the optical bench noises from the inter-spacecraft
one-way measurements $s^{\rm c}$ \cite{OttoHeinzelDanzmann}. This is
done by using the differences $\epsilon - \tau$ from each optical
bench as they contain the displacement of the optical bench relative
to the proof-mass. Second, the laser phase-fluctuations with primed
indices, $p_{i'}$, can be expressed in terms of those with unprimed
indices, $p_i$, by taking suitable linear combinations of the
$s^{\rm c}_i , \tau_i , \tau_j$.  This final operation results in the
so called $\eta$-combinations, which depend only on the three unprimed
laser phase fluctuations, $p_i$ \cite{TD2014} and are not affected by
the optical-bench noise.

As mentioned in section \ref{SECI}, recent LISA Simulation Working
Group activities have shown that the current algorithm
\cite{OttoHeinzelDanzmann} to calibrate the USO noises out of the
TDI-2 combinations is not performing as expected. To understand why,
let us consider the one-way data measurements described by equations
(\ref{s1p} - \ref{tau}). Onboard each spacecraft there are two sets of
them, one set per optical bench, with a total of $24$ data
measurements after including those for the split-interferometry
configuration and the sideband - sideband one-way inter-spacecraft
Doppler data. From a simple counting argument we conclude that the
number of observables is larger than the number of noises to be
canceled. There are $6$ lasers, three USOs, and $12$ optical-bench
noises \footnote{Although each optical bench noise is a 3-D random
  process, only its components in the plane of the array are of
  relevance.} for a total of $21$ random processes to be canceled by
properly combining the $24$ data set. To solve this well-posed
mathematical problem, the choice was made in
\cite{OttoHeinzelDanzmann} to regard one of the USO noises to be equal
to zero. This choice was based on the assumption that the USO noises
enter in the heterodyne measurements as simple differences, and that
therefore one of them could be set to zero. Unfortunately most of the
measurements do not depend on the differences of the USO noises (see
equations (\ref{s1p} - \ref{tau})). Those that do, such as the
sideband-sideband measurements, depend on differences of USO noises
measured at different times. This means that, even if all USO would
"glitch" equally, their differences would not cancel because of the
light-time delays.

As it will be shown below, there exist TDI-2 combinations that cancel
the laser phase fluctuations and from which it is in fact possible to
calibrate out the USO noises to a sufficiently high-level of
precision. This is done by properly time-shifting and linearly
combining the $\eta$-observables and the one-way carrier-to-carrier
and sideband-to-sideband measurements. 

To minimize the length of the equations we will rely on, we will use
expressions for the $\eta$-observables that display only the
contributions from the laser and USO noises. Under this assumption,
they assume the following forms
\begin{eqnarray}
\eta_{1'} & = &  \D_{2'} p_3 - p_1 + [- a_{1'} + b_{1'} ] q_1 \ ,
\label{etap}
\\
\eta_{1} & = &  \D_{3} p_2 - p_1 - a_{1} q_1 + b_2 \D_3 q_2 \ ,
\label{eta}
\end{eqnarray}
where the other four $\eta$-observables are obtained from the above
expressions by permutation of the spacecraft indices.

\section{Clock-Noise Calibration from the TDI-2 Combinations}
\label{SECIII}

We describe the procedure for calibrating the USO noises out of the
TDI-2 combinations by deriving the expressions for the unequal-arm
Michelson interferometric combination $X_1$ and the Sagnac
combination $\alpha_1$. The procedures described in this article can
easily be extended to other TDI-2 combinations and for this reason
we do not include them here.

\subsection{The $X_1$ Combination}

In terms of the $\eta$-observables, the TDI-2 unequal-arm Michelson
combination, $X^{\rm q}_1$, is given by the following expression
\cite{TD2014}
\begin{eqnarray}
X^{\rm q}_1 & = & \left[\D_{3}\D_{3'}\D_{2'}\D_{2} - I \right]
  \left[(\eta_{1'} + \D_{2'}\eta_{3}) + \D_{2'}\D_2(\eta_{1} +
    \D_3 \eta_{2'}) \right]
\nonumber \\
& & - \left[\D_{2'}\D_{2}\D_{3}\D_{3'} - I \right]
  \left[(\eta_{1} + \D_{3} \eta_{2'}) + \D_3 \D_{3'}(\eta_{1'} +
    \D_{2'} \eta_{3}) \right] \ ,
  \label{X1q}
\end{eqnarray}
where the label $^{\rm q}$ emphasizes the USO-noise dependence. After
substituting equations (\ref{etap}, \ref{eta}) in equation (\ref{X1q}) we have
\begin{eqnarray}
X^{\rm q}_1 & = & \left[\D_{3}\D_{3'}\D_{2'}\D_{2} - I \right]
\left[ b_1(\D_{2'}\D_{2} - I) q_1 - a_1 \D_{2'} \D_{2} q_1 - a_{1'} q_1 -
  a_{2'} \D_{2'}\D_{2}\D_{3} q_2 - a_3 \D_{2'} q_3 \right]
\nonumber
\\
& & - \left[\D_{2'}\D_{2}\D_{3}\D_{3'} - I \right] \left[ b_1 \D_{3}\D_{3'} 
(\D_{2'}\D_{2} - I) q_1 - a_{1'} \D_{3}\D_{3'} q_1 - a_1 q_1 - a_{2'}
\D_{3} q_2 \right.
\nonumber
\\
& & - \left. a_3 \D_{3}\D_{3'}\D_{2'} q_3 \right] \ .
\end{eqnarray}
Since the commutator of two delay operators applied to a LISA USO
noise is, to a very good approximation, equal to zero, we can rewrite
the above equation in the following form
\begin{eqnarray}
X^{\rm q}_1 & \simeq & \left[\D_{3}\D_{3'}\D_{2'}\D_{2} - I \right]
\left[ b_{1'}(I - \D_{3}\D_{3'})(I - \D_{2'}\D_{2}) q_1 + a_1 (I - \D_{2'}
\D_{2}) q_1 \right. 
\nonumber
\\
& - & \left. a_{1'} (I - \D_{3}\D_{3'}) q_1 + a_{2'} \D_{3} (I -
  \D_{2'}\D_{2}) q_2 - a_3 \D_{2'} (I - \D_{3}\D_{3'}) q_3 \right] \ ,
\label{X1qF}
\end{eqnarray}
where we have factored out the delay-operator $[\D_{3}\D_{3'}\D_{2'}\D_{2}
- I]$.

It is important to note that the $q$-terms in the square-bracket in
equation (\ref{X1qF}) can easily be related to those in equation (9)
of \cite{TEA02}, which are for the TDI-1 combination $X$ of a static
LISA. This means that the expressions calibrating the USO noises out
of the TDI-2 combination $X_1$ can be derived by using the approach of
\cite{TEA02} for the static-array unequal-arm Michelson combination
$X$.

Following Armstrong, Estabrook and Tinto \cite{TEA02}, let us
introduce the following linear combinations of the carrier-to-carrier
and sideband-to-sideband one-way heterodyne measurements
\begin{eqnarray}
r_{1'} & \equiv & \frac{s^{\rm c}_{1'} - s^{\rm sb}_{1'}}{m_3 f_3} \ ,
\label{rs1p}
\\
r_{1} & \equiv & \frac{s^{\rm c}_{1} - s^{\rm sb}_{1}}{m_{2'} f_2} \ .
\label{rs1}
\end{eqnarray}
Note that the above observables differ from those given in
\cite{TEA02} by the presence of the modulation frequency integers
$m_{2'}$ and $m_{3}$.  After substituting in equations (\ref{rs1p},
\ref{rs1}) the expressions for the one-way heterodyne measurements,
$s^{\rm c}$, $s^{\rm sb}$ (equations (\ref{s1p}, \ref{s1psb},
\ref{s1}, \ref{s1sb})) after some algebra it is possible to obtain the
following expressions for $r_{1'} \ , r_{1}$
\begin{eqnarray}
r_{1'} & = & \frac{(1 - {\dot L}_{2'})}{f_1} q_1  - \frac{\D_{2'}
  q_3}{f_3} \ ,
\label{rs1pnew}
\\
r_{1} & = & \frac{(1 - {\dot L}_{3})}{f_1} q_1  - \frac{\D_{3} q_2}{f_2} \ .
\label{rsnew}
\end{eqnarray}
By neglecting terms proportional to $\dot L$, equations (\ref{rsnew})
can be rewritten to sufficient precision in the following form
\begin{eqnarray}
r_{1'} & = & \frac{q_1}{f_1}  - \frac{\D_{2'} q_3}{f_3} \ ,
\label{rs1pf}
\\
r_{1} & = & \frac{q_1}{f_1}  - \frac{\D_{3} q_2}{f_2} \ ,
\label{rs1f}
\end{eqnarray}
with the remaining expressions obtained by cyclic permutations of the
spacecraft indices. Note that the dependence on the $m$-integers has
dropped-out from the $r$-combinations (equations \ref{rs1pf},
\ref{rs1f}), which are essentially equal to the corresponding ones in
\cite{TEA02} after modifying them to account for the inequality of the
delays experienced by laser beams propagating along opposite
directions (Sagnac effect).

Note also that, since there are only three USO noises $q_i$, there
exist relationships relating the six calibration data
$(r_{i}, r_{i'}) \ , \ i = 1, 2, 3 \ , \ i' = 1', 2', 3'$. Because the
mathematical structure of the $r$-observables (equations (\ref{rs1pf},
\ref{rs1f})) is equal to that of the one-way measurements in which the
random processes $q_i/f_i$ play the same role as the laser phase
noises, we infer that such relationships belong to the TDI-space.

By using the equations (\ref{rs1pf}, \ref{rs1f}) and following \cite{TEA02}
it is easy to derive the following identities
\begin{eqnarray}
\left[I - \D_{2'} \D_{2} \right] q_1 & = & 
f_{1} \ (r_{1'} +  \D_{2'} r_{3}) \ ,
\label{eq:14} 
\\
\left[I - \D_{3} \D_{3'} \right] q_1 & = & 
f_{1} \ (r_{1} +  \D_{3} r_{2'}) \ ,
\label{eq:15}
\\
\D_{3} \left[I - \D_{2'} \D_{2} \right] q_2 & = & 
f_2 \ [r_{1'} - (I - \D_{2'} \D_{2}) r_{1} + \D_{2'} r_{3}] \ , 
\label{eq:16}
\\
\D_{2'} \left[I - \D_{3}\D_{3'} \right] q_3 & = & 
f_3 \ [r_{1} - (I - \D_{3} \D_{3'}) r_{1'} +  \D_{3} r_{2'}] \ .
\label{eq:17}
\end{eqnarray}

By first applying the delay operator $[\D_{3}\D_{3'}\D_{2'}\D_{2} - I]$ to
equations (\ref{eq:14} - \ref{eq:17}) and substituting the resulting
expressions into equation (\ref{X1qF}) we finally find the following
USO-corrected $X_1$ combination
\begin{eqnarray}
X_1 & \equiv & X^{\rm q}_1 - \left[\D_{3}\D_{3'}\D_{2'}\D_{2} - I \right] 
\Bigl[b_{1'} \ \frac{f_1}{2} \ [(I - \D_{3}\D_{3'}) \ (r_{1'} + \D_{2'}
               r_{3}) + (I - \D_{2'}\D_{2}) \ (r_{1} + \D_{3}
               r_{2'})] \Bigr.
\nonumber
\\
& & + \Bigl. a_{1} \ f_1 \ [r_{1'} + \D_{2'} r_{3}] - a_{1'} \ f_1 \
    [r_{1} + \D_{3}r_{2'}] + a_{2'} \ f_2 \ [r_{1'} - (I - \D_{2'}\D_{2})r_{1} + \D_{2'} r_{3}] \Bigr.
\nonumber
\\
  & & 
  -  \Bigl. a_{3} \ f_3 \ [r_{1} - (I - \D_{3}\D_{3'})r_{1'} + \D_{3} r_{2'}] \Bigr]
\label{eq:18}
\end{eqnarray}
Since the unequal-arm Michelson interferometric response $X_1$ is
antisymmetric with respect to permutations of the indices (2,3'),
(3,2'), (1,1'), the corresponding combinations used for calibrating
out the USO noise from $X^{\rm q}_1$ have been antisymmetrized by
relying on the equations (\ref{eq:14} - \ref{eq:17}) and taking
advantage of the commutativity of two delay operators applied to a
LISA USO noise. The other two unequal-arm Michelson responses, $X_2$
and $X_3$, follow from equation (\ref{eq:18}) after performing a
cyclic permutation of the spacecraft indices.

\subsection{The $\alpha_1$ Combination}

In terms of the $\eta$-observables, the TDI-2 Sagnac combination,
$\alpha^{\rm q}_1$, is given by the following expression \cite{TD2014}
\begin{equation}
\alpha^{\rm q}_1  = [\D_{3}\D_{1}\D_{2} - I] \,
  [\eta_{1'} + \D_{2'} \eta_{3'} + \D_{2'} \D_{1'} \eta_{2'}] -
  [\D_{2'}\D_{1'}\D_{3'} - I] \,
  [\eta_{1} + \D_{3} \eta_{2} + \D_{3} \D_{1} \eta_{3}] \ .
\label{alpha1q}
\end{equation}
After substituting equations (\ref{etap}, \ref{eta}) in equation
(\ref{alpha1q}) we have
\begin{eqnarray}
\alpha^{\rm q}_1  & = & [\D_{3}\D_{1}\D_{2} - I] \,
  [(b_{1'} - a_{1'}) \ q_1 \ + \ \D_{2'} \D_{1'}(b_{2'} - a_{2'}) \ q_2
                        \ + \  \D_{2'} (b_{3'} - a_{3'}) q_3] 
\nonumber
\\
& & - [\D_{2'}\D_{1'}\D_{3'} - I] \,
  [(-a_1 + b_1 \D_3 \D_1 \D_2) \ q_1 \ + \ (-a_2 + b_2) \D_3 q_2 
\nonumber
\\
& & \ + \ (-a_3 + b_3) \ \D_3 \D_1 q_3] \ .
\label{alpha1qq}
\end{eqnarray}
Following a reasoning similar to the one made earlier to evaluate the
commutator of two delay-operators applied to a USO noise, it is easy
to show that also $(\D_{3'} \D_{2'} \D_{1'} - \D_3 \D_2 \D_1) q \simeq 0$, where
the $\simeq 0$ means: ``it results in relative frequency fluctuations
(strain) significantly smaller than those identified by the
acceleration and optical-path noises''.  Equation (\ref{alpha1qq}) can
therefore be rewritten in the following form
\begin{eqnarray}
\alpha^{\rm q}_1  & = & [\D_{3}\D_{1}\D_{2} - I] \,
  \Bigl[(a_{1} - a_{1'}) \ q_1 \ + \ b_{1'} (I + \D_3\D_1\D_2)q_1 \ + \ 
[\D_{2'}\D_{1'} (b_{2'} - a_{2'}) \Bigr. 
\nonumber
\\
& & \Bigl. \ + \ (a_2 - b_2)\D_3] \ q_2 \ + \   [\D_{3}\D_{1} (a_{3} - b_{3}) \ + \ (b_{3'} - a_{3'})\D_{2'}] \
    q_3 \Bigr] \ , 
\label{alpha1qqq}
\end{eqnarray}
after having factored out the delay operator
$[\D_{3}\D_{1}\D_{2} - I]$.

Since the expression inside the large square-brackets is (apart from the
primed-delays due to the Sagnac effect) the same as that of the TDI-1
combination $\alpha$ given in \cite{TEA02}, and because it was shown
there that it is impossible to exactly calibrate out of $\alpha$ the
USO noise using the $r$-combinations \cite{Hellings}, it follows that
also for $\alpha_1$ it is impossible to exactly calibrate out the USO
noise. As shown in \cite{TEA02}, however, we can rewrite the USO
phase noises in terms of some of the $r$-data and only the USO phase
noise $q_1$ by using the following additional identities
\begin{eqnarray}
\D_3 \D_1 \D_2 q_{1} & = & q_1 - f_1 \ [\D_3 \D_1 r_{3} + \D_3 r_{2} + r_{1}] \ ,
\label{eq:21}
\\
\D_{2'}\D_{1'} q_{2} & = & \frac{f_2}{f_1} \ q_1 - f_2 \ [r_{1'} + \D_{2'}r_{3'}] \ \ ,
\label{eq:24}
\\
\D_{3}\D_{1} q_{3} & = & \frac{f_3}{f_1} \ q_1 - f_3 \ [r_{1} + \D_{3}r_{2}] \ \ ,
\label{eq:25}
\\
\D_3 q_{2} & = & \frac{f_2}{f_1} \ q_{1} - f_2 \ r_{1} \ \ ,
\label{eq:34}
\\
\D_{2'} q_{3} & = & \frac{f_3}{f_1} \ q_{1} - f_3 \ r_{1'} \ .
\label{eq:35}
\end{eqnarray}

The USO noise terms involving the $q_i$ in equation (\ref{alpha1qqq}) then become
\begin{eqnarray}
& \ & 
\left[(a_{1} - a_{1'} + 2 b_{1'}) \ f_1 \ + \ (a_{2} - a_{2'} + 2 b_{2'}) \ f_2 
\ + \ (a_{3} - a_{3'} + 2 b_{3'}) \ f_3 \right] \ \frac{q_1}{f_1}
\nonumber \\
& &
- f_1 \ b_{1'} \ [r_{1} + \D_{3}r_{2} + \D_{3}\D_{1}r_{3}] - f_2 \ [b_{2'} + a_{2}]
\ r_{1} - f_3 \ [b_{3'} - a_{3'}] \ r_{1'}  
\nonumber \\
& & -
f_2 \ [b_{2'} - a_{2'}] \ [r_{1'} + \D_{2'}r_{3'}] - 
f_3 \ [b_{3'} + a_{3}] \ [r_{1} + \D_{3}r_{2}] \ .
\label{eq:26}
\end{eqnarray}
If we now take into account the expressions for the $a$-
and $b$-coefficients, the first term in equation (\ref{eq:26}) can be
reduced to the following form
\begin{equation}
\left[(\nu_2 - \nu_{3'}){\dot L_1} + (\nu_3 - \nu_{1'}){\dot L_2} +
(\nu_1 - \nu_{2'}){\dot L_3} \right] \ \frac{q_1}{f_1} \ ,
\label{eq:26bis}
\end{equation}
This corresponds to relative frequency fluctuations (or strain noise)
of the order of about $10^{-27}$ under the assumptions of having laser
frequency offsets of a few hundred megahertz, a laser center frequency
equal to $3 \times 10^{14}$ Hz, Doppler term $\dot L_i$ equal to
about $3 \times 10^{-8}$, and a USO frequency stability of
about $10^{-13}$.  Thus we can ignore it and, after some algebra,
define the laser-noise-free and USO-noise-free reduced data $\alpha_1$
to be
\begin{eqnarray}
\alpha_1 & \equiv & \alpha^{\rm q}_1 + [\D_{3}\D_{1}\D_{2} - I] \,
\Bigl[ \frac{1}{2} \ f_1 \ b_{1'} [ (r_{1} +  \D_{3} r_{2} +
                    \D_{3}\D_{1} r_{3}) 
+ (r_{1'} + \D_{2'} r_{3'} + \D_{2'}\D_{1'} r_{2'})] \Bigr.
\nonumber
\\
& & + \Bigl. f_2 (b_{2'} + a_{2}) r_{1} \ + \ f_3 (b_{3'} - a_{3'})
    r_{1'} \ + \ f_2 (b_{2'} - a_{2'}) (r_{1'} + \D_{2'} r_{3'})
    \Bigr.
\nonumber
\\
& & + \Bigl. f_3 (b_{3'} + a_{3})(r_{1} + \D_{3}r_{2}) \Bigr]
\label{eq:27}
\end{eqnarray}
Like $X_1$, also $\alpha_1$ should be antisymmetric with respect to
permutation of the indices (2,3'), (3,2'), (1,1'). The combinations in
equation (\ref{eq:26}) used for calibrating out the USO noise from
$\alpha^{\rm q}_1$ have therefore, in equation (\ref{eq:27}), been
antisymmetrized. The remaining two TDI-2 Sagnac responses, denoted
$\alpha_2$ and $\alpha_3$, follow from equation (\ref{eq:27}) after
performing cyclic permutation of the spacecraft indices.

\section{Conclusions}
\label{SECIV}

This article addresses the problem of calibrating the onboard clock
phase fluctuations out of the second-generation TDI combinations. We
have focused our analysis on deriving calibrating expressions for the
unequal-arm Michelson combination $X_1$ and the Sagnac combination
$\alpha_1$ as similar procedures can be extended to all other
second-generation TDI combinations. Our approach relies on the key
observation that the commutator of two delay operators applied to a
LISA USO noise results in relative frequency fluctuations that are
significantly smaller than those of the secondary noises and can
therefore be neglected.

Although the sideband-sideband technique suppresses the LISA's USO
noises to levels significantly smaller than that identified by the
secondary noises, it does not cancel them exactly. This might result
in a sensitivity limitation for more ambitious missions characterized
by higher sensitivities and/or significantly larger beat-notes
\cite{BBO,astrod}. This is because the calibrating algorithm presented
here might not sufficiently suppress the USO noises in their TDI-2
combinations. The use of optical frequency-comb, on the other hand,
provides a solution in these cases by generating the microwave signal
frequency coherent to the frequency of the onboard laser. This is
because the optical frequency-comb technique exactly cancels the
microwave signal phase fluctuations by using modified TDI-2
combinations and it does not require modulated beams. 

In addition, use of an optical frequency comb may result in a
simplification of the LISA's onboard interferometry system
\cite{TintoYu2015} because (i) generation of modulated beams and
additional heterodyne measurements involving clock microwave sidebands
will no longer be needed, and (ii) the entire onboard USO subsystem
can be replaced with the microwave signal referenced to the onboard
laser. This may result in a reduced system complexity and probability
of subsystem failure. Recent progress in the realization of a
space-qualified optical frequency comb indicates that such a capability
will be available well before LISA's launching date \cite{Holzwarth}.

\section*{Acknowledgments}

M.T. thanks Dr. Frank B. Estabrook and Dr. John W. Armstrong for their
constant encouragement during the development of this
work. O.H. gratefully acknowledges support from the Deutsches Zentrum
f\"ur Luft- und Raumfahrt (DLR) with funding from the
Bundesministerium f\"ur Wirtschaft und Technologie (Project
Ref. No. 50OQ1601).

\bibliography{References}
\end{document}